\documentstyle[preprint,version2,aps,12pt]{revtex}
\begin{document}

\newcommand{\beq}{ \begin{equation} }
\newcommand{\eeq}{ \end{equation} }
\newcommand{\bea}{ \begin{eqnarray} }
\newcommand{\eea}{ \end{eqnarray} }

\draft
\preprint{IFUP-TH 42/94 \ \ \ \ \ \ }
\vspace{1cm}
\begin{title}
{\bf Heating and small-size instantons in the $O(3) \: \sigma $ model \\ 
on the lattice}
\end{title}
\vspace{1cm}
\author{Federico Farchioni and Alessandro Papa}
\vspace{0.4cm}
\begin{instit}
Dipartimento di Fisica dell'Universit\`a and I.N.F.N. \\
Piazza Torricelli 2, I-56126 Pisa, Italy.
\end{instit}
\vspace{1cm}
\begin{abstract}
We study the role of small-size instantons in the determination of the
topological susceptibility of the 2-d $O(3) \: \sigma $ model on the lattice.
In particular, we analyze how they affect the non-perturbative determination, 
by Monte Carlo techniques, of the renormalizations on the lattice. 
As a result, we obtain a high-precision non-perturbative determination of  
the mixing with the unity operator, finding good agreement with perturbative
computations. We also obtain the size distribution of instantons 
in the physical vacuum up to very 
small values of the size in physical units, without observing 
any ultraviolet
cut-off. Moreover, we  show by analytical calculation
that the mixing of the topological susceptibility with the action density is 
a negligible part of the whole non-perturbative signal.
\end{abstract}
\newpage

\narrowtext
\section{\bf Introduction}
\label{sec:intro}
\pagestyle{plain}

\pagenumbering{arabic}

Simulating a
theory on the lattice is the only reliable instrument to study its  
non-perturbative aspects: the lattice theory, merely being  an UV regularized
version of the theory on the continuum, takes into account all non-perturbative
fluctuations. 
Some quantities of physical interest,  which can be extracted 
from the lattice,
are related to vacuum expectation values of composite operators.
Lattice Monte Carlo simulations give a numerical estimate 
of the cut-off-dependent
{\em bare} expectation values, while the physical quantities on the continuum 
are the 
{\em renormalized} expectation values, which are cut-off-independent. So, the 
physical quantities  can be determined from the lattice only if 
the renormalizations of the lattice-regularized operators are
completely under control.

Perturbation theory has been, up to now, the only mean for the evaluation
of the lattice renormalization constants. In particular, perturbative
techniques, combined with Monte Carlo simulations, have been  
applied to the problem of the determination from the lattice of the 
topological susceptibility of the $QCD$ vacuum \cite{CDP}\cite{CDPV}.
The situation has recently changed, since a new  method for the
determination of the renormalizations of the lattice
topological susceptibility has been found \cite{Teper1}\cite{DG-V}. 
This method, known in literature as the ``heating"  method, is fully 
non-perturbative since it relies only on Monte Carlo techniques. Applications  
to $CP^{N-1}$ models \cite{CRV2} and  $SU(2)$ Yang-Mills theory
\cite{ACDGV} have shown a clear agreement with perturbative calculations  
\cite{ACDGV}\cite{noio}. In 
$SU(2)$ it has  been possible to obtain also an indirect determination of
the gluon condensate, finding agreement with previous standard Monte Carlo 
determinations \cite{gluon}.

The $O(3) \: \sigma $ model is the simplest model displaying a non-trivial 
topology, and so  it appears as  the best laboratory to test
the heating method. The drawback is that this topology is suspected to be
pathological: the semiclassical approximation 
\cite{semicl} shows  a 
small-size divergence in the instantons contribution to the partition function.
Such an ultraviolet dominated topology is expected to strongly affect
the heating method, which relies on a decoupling between short-range
perturbative modes and topological fluctuations, assumed to be long ranged. 

In fact, the method had its first application just in 
$O(3) \: \sigma $ model \cite{DG-V} \cite{noi},
but the statistical fluctuations in the numerical results of those first works
are likely to mask an eventual exotic behavior of the model;
so, a more accurate investigation on the matter is asked for.

In the present work we perform a careful analysis of the heating method
in the $O(3) \: \sigma $ model to check if, and to what extent, the 
ultraviolet feature of the topology spoils the non-perturbative 
determinations of the renormalizations of the lattice susceptibility.

\section{\bf The model}

The 2-d $O(3) \: \sigma$ model or $CP^{1}$ model  is described by the 
Lagrangian:

\beq
{\cal L} = \frac{\beta}{2} \: \partial _{\mu} \phi(x) \cdot \partial _{\mu} 
\phi(x) \;\;\;\; ,
\label{eq:def}
\eeq
where $\phi (x)$ is a three component real field satisfying the constraint
$\phi \cdot \phi = 1$.

The 2-d $O(3) \: \sigma$ model plays an 
important role in quantum field theory because it resembles in several aspects
the 4-d non-Abelian gauge theories: asymptotic freedom,  
non-perturbative behavior in the infrared region with spontaneous mass 
generation, non-trivial topological structure.

The topological charge of a spin field $\phi (x)$, $Q$, is the number of times 
$\phi (x)$ winds the sphere $S^{2}$. It can be expressed as the integral over
the space-time of a local operator, $Q(x)$:

\beq 
Q(x) = \frac{1}{8\pi} \epsilon_{\mu \nu }
\epsilon_{ijk} \phi_{i} (x) \partial _{\mu} \phi_{j}(x) \partial _{\nu} 
\phi_{k}(x)  \;\;\;\;\;  ;
\eeq
$Q(x)$ is the divergence of a topological current $K_{\mu}$ 
\cite{hof:cur}\cite{dad:cur},
\beq
Q(x)=\partial _{\mu }K_{\mu }(x) \;\;\; .
\eeq

All the classical 
solutions with non-trivial topology, the $k$-instantons, have been explicitily 
found \cite{bp}.
At a quantum level, the only available prediction comes from
the semiclassical approximation, which gives for the size distribution
of instantons in the physical vacuum:
\beq
\frac{dN}{Vd\rho} \; = \; e^{-4\pi \beta}\beta^2\frac{f(\rho M)}
{\rho^3} \; ,
\label{eq:size}
\eeq
where $M$ is the cutoff mass; renormalization group theory implies 
\beq
f \propto (\rho M)^{4\pi\beta_0}, 
\label{eq:size1}
\eeq
where $\beta_0 \; = \; \frac{1}{2\pi}\; $ is the one-loop
coefficient of the $\beta$ function; as a result: $\; dN/d\rho \propto 1/\rho$.

The generalization for $CP^{N-1}$ is:$\; dN/d\rho \propto \rho^{N-3}$.
For $SU(N)$, Eq.~(\ref{eq:size}) and Eq.~(\ref{eq:size1}) become
respectively
\beq
\frac{dN}{Vd\rho} \; = \; e^{-8\pi^2 /g^2}g^{-8}\frac{f(\rho M)}
{\rho^5} \; 
\eeq
and
\beq
f \propto (\rho M)^{16\pi^2\beta_0} \; ; 
\eeq
it follows: $\; dN/d\rho \propto \rho^{\frac{11}{3}N - 5}$.
For both $CP^{N-1}$, with $N \geq 4$, and $SU(N)$, the size distribution is 
suppressed at small sizes with a power law, while in the $O(3) \: \sigma $ 
it diverges. This makes
$O(3) \: \sigma $ model singular, and a behavior radically different
from the theory of physical interest, $QCD$, is expected at small distances.

\section{\bf The field theoretical method}

We regularize the theory on the lattice by taking the Symanzik tree-level
improved action \cite{syman}
\beq
S^{L}=-\beta \sum_{x,\mu}\: \left [ \frac{4}{3}\phi (x) \cdot \phi (x+\mu )
- \frac{1}{12} \phi (x) \cdot \phi (x+2\mu) \right ] \;\; .
\label{eq:slatt}
\eeq

In the field theoretical method, a lattice topological charge density operator
is defined as a local operator having the appropriate classical continuum limit
\cite{div:lim}; our choice is \cite{noi}:
\beq
Q^{L} \: = \: \frac{1}{32\pi } \epsilon _{\mu \nu }\epsilon _{ijk} \phi _{i}(x)
\left(\phi _{j}(x+\mu ) \: - \: \phi _{j}(x-\mu ))(\phi _{k}(x+\nu ) \: - \: \phi _{k}(x-\nu )\right) .
\eeq

It has been shown in the framework of perturbation theory  \cite{noi} that
the operator $Q(x)$ defined on the continuum is invariant under the 
renormalization group.
 
A finite multiplicative renormalization connects the matrix elements of 
$Q^{L}(x)$ with those of $Q(x)$ defined on the continuum. The 
antisymmetry of $Q^{L}(x)$ forbids mixings with any other $O(3)$
invariant operator of dimension two or less. So, the connection 
with the continuum is:
\beq
Q^{L}(x) \: = \: a^2 \, Z(\beta) \: Q(x) \: + \: O(a^4) \; \; .
\label{eq:zq}
\eeq

The topological susceptibility is defined on the continuum 
as the correlation at zero momentum of two
topological charge density operators, $Q(x)$~:
\beq
\chi \: = \: \int \: d^{2}x \;\langle 0|\:T\, [\:Q(x)Q(0)\:]\: |0 \rangle \:\:\: .
\label{eq:chi}
\eeq

The prescription defining the product of 
operators in Eq.~(\ref{eq:chi}) is \cite{cre:ope}
\beq
\langle 0|\: T [\:Q(x)Q(0)\:]\: |0\rangle\: \equiv \partial _{\mu }\langle 0|\: T[\:K_{\mu }(x)Q(0)\:]\:  |0\rangle \: \: .
\label{eq:cimento}
\eeq
This prescription eliminates the contribution of possible contact terms 
(i.e. terms proportional to the $\delta $ function or its derivatives) when
$x\rightarrow 0$.

The lattice-regularized version of $\chi$ is 
\beq
\chi ^{L} \: = \: \left\langle \;\sum_{x} Q^{L}(x)Q^{L}(0)\; 
\right\rangle \: = \: \frac{1}{L^2}\left\langle ( \:\sum_{x} Q^{L}(x) 
\:)^{2} \right\rangle \: \:\; ,
\label{eq:susc}
\eeq
where $L$ is the lattice size.

A prescription equivalent to 
Eq.~(\ref{eq:cimento}) does not exist on the lattice, and therefore the
contribution of the contact terms must be isolated and subtracted. These
contact terms appear as mixings with the action density $S(x)$ and with
the unity operator $I$, which are the only available operators with 
equal dimension or lower. In formulae
\beq
\chi ^{L}(\beta ) \: = \: a^{2}\:Z(\beta )^{2}\,\chi \: + \: a^{2}\:A(\beta )\,\langle S(x) \rangle \: + \: P(\beta ) \,\langle I \rangle \: + \: O(a^{4}) \: \: ,
\label{eq:ope}
\eeq 
where $a$ is the lattice spacing.
In Eq.~(\ref{eq:ope}) the quantity
$\langle S(x) \rangle$ is intended to be the non-perturbative part of
the expectation value of the action density, i.e. it is a signal of
dimension two.

$Z(\beta)$ and $P(\beta)$ have been calculated 
perturbatively in Ref. \cite{noi} up to the order $1/\beta^2$
and $1/\beta^5$, respectively. 

A relation similar to Eq.~(\ref{eq:ope}) holds in $SU(2)$ 
with the appropriate powers of $a$ and 
$\langle S(x)\rangle \rightarrow \langle g^2/(4\pi)^2\, 
F^{a}_{\mu\nu}F^{a}_{\mu\nu}(x) \rangle$,   
the gluon condensate. In $SU(2)$, the mixing with the gluon condensate
is an appreciable portion of the whole non-perturbative signal in the scaling
region, and can be detected through $\chi^L$ \cite{ACDGV}.

We realize that the situation is radically different in the case of the 
$O(3) \: \sigma $ model. Indeed, calculating the 
leading perturbative contribution to the 
mixing coefficient, and using a large $N$ result for $\langle S(x) \rangle$ 
\cite{roca}
(see Appendix for the details of the calculation and the discussion), we find
that the mixing term is smaller than one thousandth of the total
non-perturbative signal in the scaling region observed in Ref. \cite{noi}. We
therefore argue that the contribution to $\chi^{L}$ coming from the  mixing 
term can be safely neglected.

\section{\bf Numerical results}

\subsection{\bf Heating and cooling}

Now we  want to give a brief account of the heating method (for Ref.,
see \cite{Teper1}\cite{DG-V}). This method allows a non-perturbative 
determination of the lattice renormalization constants, $Z(\beta)$ and 
$P(\beta)$, which have their origin in the short-range fluctuations 
$(l \sim a)$.
 
Ensembles  of configurations $\{C_t\}$ are constructed on the lattice,
 each configuration of the ensemble being obtained 
by performing  a sequence of $t$  local Monte Carlo sweeps starting from
a discretized classical configuration $C_0$ - 
a large instanton or the flat configuration. 
For small $t$, the heating process does nothing but
``switch on" the small-range fluctuations which are responsible for the 
renormalizations: when the starting configuration is a large instanton
(flat configuration), 
measuring $Q^{L}$$(\chi^L$) on the ensembles $\{C_t\}$, a plateau at the value 
of  $Z(\beta)$ $\,(P(\beta))$ is expected after a certain time, 
not depending on
$\beta$, corresponding to the time of thermalization of quantum fluctuations.

If the heating is protracted, the (local) algorithm of thermalization 
generates fluctuations of 
ever increasing size according to the random walk law $l^2 \propto t$. 

In the case
of the heating of the flat configuration, when the ranges $l \sim \xi_{eq}$ 
- the equilibrium correlation length - are reached, the contribution of 
 the mixing with 
the  action density to $\chi^L$ is thermalized. This is clearly observed 
in $SU(2)$, where a plateau at a value of $\chi^L$ exceeding
$P(\beta)$  (perturbatively calculated)
of the amount of the expected mixing, is detected
\cite{ACDGV}. The display of this plateau is possible in $SU(2)$ since the
average number of instantons in the first stage of the heating is nearly zero
- instantons exhibit a very severe form of critical slowing down.  
Otherwise, they would start to give a contribution to the topological 
non-perturbative part of $\chi^L$ before the thermalization of the mixing with
the action density  is reached, so preventing the observation of the plateau.
 
It is  interesting to investigate if this favourable situation 
happens also in the case of the $O(3) \: \sigma$ model; here, the 
small-size divergence in the instanton size distribution  is likely to cause
a radical change of the scenario. 
Since the mixing term is negligible, a plateau is expected at 
a value of $\chi^{L}$ corresponding to $P(\beta)$. 
Previous results \cite{DG-V}\cite{noi} seem
to be in agreement with this expectation; however, the statistical
fluctations of those determinations prevent to absolutely exclude 
the eventual  onset of instanton contribution during the first phase of 
the heating. 

In order to  unmask the topological structure of the configurations
in the heated ensemble, we use the ``cooling" procedure \cite{Teper2}. 
It consists in a local minimization of the action with  the purpose of 
destroying the quantum fluctuations, trying to preserve 
the background topological structure. 
We use an unconstrained cooling, consisting in the following replacement:
\beq
\phi (x) \: \rightarrow \: \phi^{\prime} (x) = \alpha\sum_{\pm\mu}\left 
[ \frac{4}{3}\phi (x+\mu) \: -  \: \frac{1}{12} \phi (x+2\mu) \right ] \; ,
\eeq
where $\alpha$ ensures the normalization of the new spin; this  
replacement exactly minimizes the action when the other spins are kept
fixed. We measure the topological charge after each step of cooling.
Typically, two behaviors are observed: the charge can rapidly go 
to zero, so revealing the absence of a topological background; otherwise,
it reaches, after a short time, a constant value next to an integer, so
indicating the presence of background instantons.

The irrelevant terms in the lattice action adopted 
- the Symanzik tree-level improved action - make the lattice action 
calculated on an instanton decrease
when its size decreases. As a consequence,  an instanton experiences 
under cooling  a progressive shrinking, 
up to its destruction. The conclusion is that the cooling procedure affects
in a certain degree the background topology, cutting off instantons with
size smaller than a certain size, depending on the number of cooling steps 
performed. This number must be big enough to clear off the quantum noise,
but not so much to completely erase most of the topological configurations.

\subsection{\bf Heating an instanton}

Here, we apply the method suggested by Teper \cite{Teper1} and already
realized in \cite{DG-V}\cite{noi}, for the determination of $Z(\beta)$. Our
contribution is an improvement of the statistic 
with the purpose to reveal eventual 
discrepancies from the two-loops perturbative 
calculation; moreover, we study of the onset of small instantons
over the starting  topological background. 

We put an instanton of charge $Q_0 = 1$ and size $20$ 
lattice units, in the middle of a $120 \times 120$ lattice. Starting from
this configuration, we construct the ensembles $\{C_{t}\}$ 
performing $t$  sweeps of a standard heat-bath algorithm, 
 and measure the average
topological charge over each ensemble $\{C_{t}\}$. 

In Fig.~2 we present
the results for three different values of $\beta $. 
We observe that a plateau is reached after about $11$ heating steps. The
independence from $\beta$  of the starting point of the plateaux has been
observed also in previous works: it reveals that  $Z(\beta)$ takes its origin
from fluctuations of small size, whose thermalization does not undergo
critical slowing down.

In Table II we compare $Z(\beta)$ 
calculated at two loops with the non-perturbative estimate obtained 
by fitting values at the plateaux. We attribute the small discrepancy 
to further terms of the perturbative expansion of $Z(\beta)$: a fit gives
$z_3 \, = \, 0.097(8)$ and $ z_4 \, = \, -0.422(12)$, 
$\, \chi^2/$ d.o.f. $\, = \, 0.15$.

In order to check the extent of  small instantons 
production during heating, we have
analyzed a sample of $1000$ configurations obtained after
$15$ heating steps at $\beta = 1.45$, where the small instantons contamination
is expected to be the largest among the values of $\beta$ considered. 
We classify each 
configuration assigning it to a definite
topological sector according to its cooled charge.
We find that $\, \sim30$ configurations have left the $Q = 1$ sector: 
$\, \sim20$ migrating into  the zero  
charge sector, and $\, \sim10$ into the $Q = 2$ sector.
This effect is to be attributed to  the generation 
of small-size instantons (of charge $-1$ and $+1$ respectively),
laying on the background topological configuration. 
We explain the  small asymmetry  observed as the effect of the explicit
breaking of the charge symmetry by the starting configuration: 
starting from the $Q= 1$ sector, it is 
energetically more convenient to fall into the $Q=0$ sector than to rise
into the $Q=2$ sector.
We have verified that the systematic error induced on the determination
of $Z(\beta )$ by a  topological contamination of the observed
 extent
is negligible within the statistical uncertainity of our result.

\subsection{\bf Heating the flat configuration}
Here, in addition to the standard
heat-bath algorithm, we use a faster local algorithm of thermalization, 
the overheat-bath 
\cite{noi}.

First of all, we have analyzed the thermalization process
during heating. We have  observed the behavior under heating of $\xi_{\perp}$,
the correlation length of the components of the spins orthogonal to
the direction of polarization of the starting flat configuration; 
with this choice we get rid of the effects of $O(3)$ symmetry breaking of the
starting configuration.

We have used the following definition \cite{CRV1}:
\beq
\xi_{\perp}^2 \: = \: \frac{1}{4\sin^2(\pi /L)}\left [ \frac{\tilde{G}_{\perp}
(0,0)}{\tilde{G}_{\perp}(0,1)}\, - \, 1 \right ] \;\; ,
\label{eq:xiort}
\eeq
where 
\beq
\tilde{G}_{\perp}(k) \: = \: \frac{1}{L^2}\sum_{x,y}\langle \phi_{\perp}(x)
\cdot\phi_{\perp}(y) \rangle \exp\left [ i\frac{2\pi}{L}(x-y)\cdot k \right ]
\; \; .
\eeq
We have verified that this quantity  reproduces at thermal equilibrium
the value $\xi_{eq}$ obtained with the standard definition. 
In Figs.~3, 4 we present some values of $\xi_{\perp}^2$ versus $t$, 
the number of heating  steps performed, obtained with the two algorithms
of thermalization.
We observe that the expected law $\; \xi_{\perp}^2(t) = c\, t\; $ 
fits data nicely, with $c_{h} = 0.335\pm 0.042$, $\chi^2$/d.o.f.~$\sim0.01$,
for the heat-bath, 
and $c_{ov} = 1.89\pm 0.16 $, $\chi^2$/d.o.f.~$\sim0.79$, for the overheat-bath. 
We note that the overheat-bath is about six times faster than the heat-bath.

In $SU(2)$ Yang-Mills theory,  topological fluctuations
are decoupled  from the ordinary fluctuations as for thermalization time.
This property allows the observation,
during the heating of $\chi^{L}$ starting from the flat configuration,
of an extended plateau corresponding to the non-topological
contribution to $\chi^{L}$, i.e. the mixing with the identity operator 
plus the mixing with the gluon condensate \cite{ACDGV}. Should this hold also
in the $O(3) \: \sigma$ model, a plateau in correspondence to $P(\beta)$
would be detectable, and, as a result, a non-perturbative evaluation
of such quantity could be obtained - the mixing term is here negligible.

We have performed, at various values of $\beta$, measures of $\chi^L$ 
during the heating of the flat configuration with the slower algorithm
of thermalization,
the standard heat-bath, each ensemble $\{C_t\}$ containing from $10000$ to 
$40000$ configurations. Unfortunately, we could not clearly single out 
any plateau, always obtaining a steady drift to the equilibrium value 
(see Fig.~5).  The situation is the same when the overheat-bath is used, with
a six scale factor on the times of thermalization.

Being aware of the ultraviolet nature of the model in study,
the explanation of this behavior is straightforward:
small-size instantons are precociously generated by the heating algorithm. 

In order to check this diagnosis,
we sistematically cool  ensembles $\{C_t\}$, 
obtained with a standard heat-bath, at values of $t$
beyond  the thermalization time of the perturbative
fluctuations indicated by the previous works on this matter
\cite{DG-V}\cite{noi}. 
We find that five steps of uncostrained cooling are 
sufficient to unmask  the topological content of the configurations 
through the observation of the behavior of the topological charge.
If cooling is further protracted a part of instantons - the smaller ones -
are destroyed: this phenomenon is revealed by the behavior of the charge 
under cooling that, after a short plateau at a value near to an integer,
goes to zero in few steps.
The result is that, as expected, the presence of topological fluctuations is 
relevant already at the first steps of the heating process:  
the argument that instantons are  subject to a special 
kind of critical  slowing down, more severe than non-topological fluctuations,
seems not to apply to this model.  

Now, if topological configurations are sistematically removed
from the statistical ensembles $\{C_t\}$, and $\chi^{L}$ is averaged
only over the trivial configurations, a long plateau should be observed
after the thermalization of the perturbative modes.
This plateau would correspond {\em just} to $P(\beta)$ as defined in 
Eq.~(\ref{eq:ope}).

In order  to get rid of instantons, we 
sistematically discard all configurations having, after $5$ steps of
unconstrained cooling, a topological charge greater than a fixed threshold 
value. An analysis at $\beta = 1.45$ on a sample $\{C_{70}\}$ consisting of 
$100$ configurations 
shows that this procedure eliminates almost all configurations with
non-trivial topology. In Fig.~6
we show the results obtained at $\beta = 1.45$ for a chosen threshold: 
a clear long  plateau is now observed. We have checked that data 
are stable under a small change of  the threshold: this ensures that 
the plateau observed  is a real effect and not an artifact of the 
subtraction procedure. 
We have repeated the above procedure for other values of $\beta$ : 
a plateau is observed for $t \geq \bar{t} \simeq 35$,
not depending on $\beta$ as expected; $\bar{t}$ is the time of 
thermalization of fluctuations with size $l \simeq  4$ lattice spacings
which, within our error bars, saturate the perturbative signal, $P(\beta)$.
We observe that $P(\beta)$, in comparison with $Z(\beta)$, is sensitive to 
larger size fluctuations: an indication of this could already be taken out from
the perturbative series of $P(\beta)$, which starts to get contribution from 
higher orders in $1/\beta$.  
In Table III  
values of $P(\beta)$ fitted on the plateaux, $P_{np}(\beta)$, are compared 
with those  of $P_{pert}(\beta)$. The latter quantity  is obtained 
in Ref.~\cite{noi} by a perturbative calculation combined with a fit on 
$\chi ^{L} (\beta)$ at large $\beta$: in this region, the probability of 
occurrence of instantons is exponentially suppressed, 
as well  the non-perturbative contribution of other
origins, so  $P_{pert}(\beta)$ 
is in all equivalent to a truncated perturbative determination of $P(\beta)$.
The agreement  observed in Table III reveals
that $P(\beta)$ is a substantially perturbative quantity,   
well approximated by the first few perturbative terms. This result is not
trivial, since in $O(N) \: \sigma$ models perturbative series are not 
Borel-summable~\cite{David}.

Correlations between measures have been taken into account 
performing standard binning procedures; we observe in this regard that 
instantons, which are the main source of correlations
for $\chi^{L}$, have been rejected from the ensemble, and so
the problem of correlations is less critical than usual.

In order to study the approach to thermal equilibrium of $\chi^L$ during the 
heating of the flat configuration, we exploit the faster overheat-bath 
algorithm. In Fig.~7 we show  the behavior of $\chi^L$ during the heating
at $\beta = 1.45$: the time required to reach the equilibrium value is
consistent with $t_{eq} = \xi^2_{eq}/c_{ov}$.
This observation is confirmed, within our error bars, for
other values of $\beta$. The suggestion we get from this evidence is 
that instantons undergo the ordinary slowing down: the time of 
thermalization of all fluctuations depends only on their size - 
no matter of their topology.

We further observe that the time  $\chi^L$ takes to overcome 
its perturbative value (see Fig.~8 and Table III) is the same for different 
values of $\beta$. By a measure at $\beta = 2.50$, we have verified that this 
is just the time the algorithm takes to thermalize the perturbative 
fluctuations. The observed time ($\bar{t} \simeq 8$ heating steps)
corresponds, according to the random walk law, to thermalized lengths 
of $\sim 4$ lattice spacings, in agreement
with the results obtained with the heat-bath algorithm. The 
suggestion is that there is no non-perturbative contribution to $\chi^L$
before perturbative modes are thermalized; otherwise, such non-perturbative 
contribution would give an additional boost to the signal, making it to
overcome the perturbative threshold in advance at small $\beta$ 
in comparison with 
the time taken at $\beta = 2.50$ to saturate the perturbative signal.

\subsection{\bf Small instantons size distribution}

On the basis of the results of the previous subsection, we make the following 
work hypothesis about the process of thermalization of topological 
fluctuations: we assume that the heating algorithm 
reproduces on the lattice at the time $t$ the equilibrium size 
distribution of the average instanton number $dN/d\rho$, up to the length
thermalized at that $t$, $l(t)$; we moreover assume that the  quantity
$\xi_{\perp}(t)$, defined in (\ref{eq:xiort}), is a measure of $l(t)$.
In this hypothesis, the thermalized length obeys the law
$l(t)\, = \,\sqrt{c\, t}$, where $c$ is the same constant
that enters the random walk law for $\xi_{\perp}$ and that has been
estimated for the two algorithms of thermalization in the previous subsection.
Now, called $N(t)$ the average number of instantons in the 
ensemble at the time $t$, we write
\beq
N(t) \: = \: \int ^{l (t)}d\rho\;  \frac{dN}{d\rho}  \; \; .
\label{eq:n}
\eeq

In the hypothesis of non-interacting instantons, the average number of 
instantons in the statistical ensemble, $N$, is simply related to the
topological susceptibility: $\chi = N/V$.
Indeed, denoting with $n(\bar{n})$ the number of (anti-)instantons, 
and with $P_{n,\bar{n}}$ the probability to have a configuration 
with $n$ instantons and $\bar{n}$ anti-instantons, the following relations
hold
\beq
\chi \: \equiv \: \frac{1}{V}\sum_{n,\bar{n}}(n-\bar{n})^2 P_{n,\bar{n}} \: = 
\: \frac{1}{V}\sum_{n,\bar{n}} (n + \bar{n}) P_{n,\bar{n}} \: \equiv \: 
\frac{N}{V} \; \; .
\label{eq:enne}
\eeq
The hypothesis of non-interacting instantons plays a role in
the second equality of (\ref{eq:enne}) since, in this case,  
\beq
P_{n,\bar{n}} \: = \: \frac{1}{2^{n+\bar{n}}}\frac{(n+\bar{n})!}{n!\; \bar{n}!}
P_{n+\bar{n}} 
\eeq
($P_N$ is the probability of a configuration with $N$ topological
objects, instantons or anti-instantons). We note, by the way, that 
the above condition is satisfied if $P_{n,\bar{n}} = P_{n}\: P_{\bar{n}}$
with $P_{n(\bar{n})}$ poissonian, as in the semiclassical approximation.

Eq.~(\ref{eq:enne}), when translated on the lattice in the framework of 
the field theoretical method,  allows to obtain 
a measure of the average number of
instantons in the ensemble $\{C_t\}$, $N(t)$, from the 
measure of $\chi^{L}$:
\beq
\frac{N(t)}{L^2} \: = \: \left.\frac{\chi^L(\beta) - P(\beta)}{Z^2(\beta)}
\right |_{t} \:
\equiv \: \left.\chi^{L}_{np}\right|_{t} \; \; .
\label{eq:nchi}
\eeq
So, we are in a position to check our hypotesis (\ref{eq:n}):
in Fig.~9 we show $N_{\chi}(t)$ as obtained from 
(\ref{eq:nchi})
at $\beta = 1.45$ together with the curve obtained from 
(\ref{eq:n}) with the choice $dN/d\rho = A/\rho ^{1.65}$: we observe 
that the behavior of $N(t)$ is nicely reproduced. 

The average number of instantons in the ensemble 
at time $t$ can be directly obtained - without passing through
$\chi^{L}$ - from the analysis of the frequency of occurrence of 
configurations in the various topological sectors. Indeed, from the
hypothesis of non-interacting instantons, it follows (see (\ref{eq:enne})):  
\beq
N \: = \: \sum_Q Q^2 P_Q  \; \; .
\label{eq:nq}
\eeq
In order to assign each 
configuration  to its  topological sector $Q$, we exploit the 
above described cooling procedure.

In Fig.~10 we show the behavior of $N_Q(t)$ so obtained
at $\beta = 1.45$. We see that $N_Q(t)$ is $\sim 0$ for $t \leq
2$ heating steps, corresponding to instantons of size $\sim2$ lattice 
spacings: instantons smaller than this size do not survive the cooling 
process. \newline
For $2 \leq t \leq \bar{t}$ ($\bar{t}$ is the time of thermalization of 
perturbative fluctuations, corresponding to $\sim4$ lattice spacings), 
$N_Q(t)$ is
different from zero even if $\chi^{L}$  has yet no non-perturbative 
contribution, as observed in the previous subsection. We conclude that 
instantons with size lower than $\sim4$
lattice spacings - though present in the ensembles $\{C_t\}$ - give no
non-perturbative contribution to $\chi^{L}$, so having the same effect 
of perturbative fluctuations. \newline
For $t \geq \bar{t}$, $N_Q(t)$ is systematically  greater 
than $N_{\chi}(t)$: this discrepancy is simply explained, since
$\chi^{L}$ loses the part of the topological contribution coming
from instantons with sizes ranging from $\sim2$ to $\sim4$  lattice spacings.
In formulae:   
\beq
N_{\chi}(t) \: = \: \int^{l (t)} _{\sim 4}d\rho\;  
\frac{dN}{d\rho}\; , 
\eeq
while
\beq
N_Q(t) \: = \: \int _{\sim 2} ^{l (t)}d\rho\;  \frac{dN}{d\rho}\; . 
\eeq
The mismatch is expected to be just 
\beq
\int _{\sim 2} ^{\sim 4}d\rho\;  \frac{dN}{d\rho}\, \simeq \, N_Q(\bar{t})\; ,
\label{eq:pre}
\eeq
for all $t\geq\bar{t}$. We find that the quantity 
$N_Q(t)-N_{\chi}(t)$ is well fitted by a constant
($\chi^2$/d.o.f $\, \sim0.1$) whose  value is
consistent with the prevision (\ref{eq:pre}) -  see Fig.~11. 
The same happens at $\beta = 1.525$. 

On the basis of the assumption (\ref{eq:n}) on the mechanism of thermalization
of instantons during heating, we can extract from the lattice the size 
distribution of instantons in the physical vacuum, $dN/d\rho$.
Indeed, from Eq.~(\ref{eq:n}) it follows   
\beq
\left.\frac{dN}{d\rho}\right|_{\rho\: =\: l(t)} 
\: = \: \frac{1}{dl(t)/dt}\left (\frac{dN}{dt}\right )\; \; .
\label{eq:dist}
\eeq
In Fig.~12 we plot $dN_{Q}/d\rho$ and $dN_{\chi}/d\rho$ versus $\rho$
at $\beta\, =\, 1.45$ as obtained from Eq.~(\ref{eq:dist}).
We find that the two determinations of $dN/d\rho$ nicely overlap. The same is 
done in Fig.~13 at $\beta = 1.525$; data  at $\beta = 1.525$ have a 
suppression $\propto a^3$ in comparison with data at $\beta = 1.45$, since
$dN/Vd\rho$ is a renormalization group invariant quantity when $V$ and 
$\rho$ are measured in physical units.

In Fig.~14 we collect all our data for $dN/Vd\rho$ 
expressing all quantities in physical units. Data coming from
different values of $\beta $,  covering a spectrum of $\rho$ ranging from 0.2 
to 0.9  units of the inverse mass gap, dispose within statistical 
fluctuations along the same curve, as expected.
A global fit with a function $A/\rho ^n$ shows that $n$ from 1 to 2
is consistent with data. In fact, $n$ seems to increase with $\rho$: 
initial data give $n \simeq 1.5$, while data at $\rho \geq 0.4$ give 
$n\simeq 2$. This behavior is intermediate to the 
expectation $n = 1$, from semiclassical prediction at small 
sizes, and the result $n = 3$ recently obtained in \cite{ingl}, where 
$dN/Vd\rho$ is determined with different techniques in the region $1 \leq
\rho \leq 5.5$.  These results seem to 
suggest for $dN/Vd\rho$ an ever more severe suppression as $\rho$ increases.

Up to the sizes investigated by our analysis 
($\rho \simeq 0.2$ inverse mass gap units), no physical ultraviolet cut-off 
has been observed in the instanton size distribution.  Due to this ultraviolet
dominance, $\chi^{L}_{np}$ loses a noticeable part of the 
non-perturbative contribution, if the discretization is 
too coarse: we argue that this is in fact  the case 
for the values of $\beta$ of the numerical simulations up to now performed. 
Since the amount of the lost contribution decreases as $\beta$ increases, 
a scaling defect for  $\chi^{L}_{np}$ should be observed. In the following, 
we show the results of $\chi^{L}_{np}/f^2(\beta)$ at thermal equilibrium
for three values of $\beta$ in the expected scaling region; 
$\; f(\beta)=2\pi\beta e^{-2\pi\beta}$ is the two-loop 
renormalization group function and
$\chi^{L}_{np}$ has been obtained from standard Monte Carlo simulations
using the non-perturbative determinations of $P(\beta)$ and $Z(\beta)$:

\bea
\beta = 1.45 \; & : & \; \; \; \; \; \; 196 \pm 11  \\ 
\nonumber 
\beta = 1.50 \; & : & \; \; \; \; \; \; 224 \pm 13  \\ 
\beta = 1.525 \; & : & \; \; \; \; \; \; 250 \pm 14 \;\;  . 
\nonumber
\eea

The increasing behavior observed indicates that,
for the values of $\beta$ in study,  $\chi^{L}_{np}$ has not yet 
saturated all topological contributions.

\section{\bf Conclusions}

In this paper we have studied the behavior of topological small-size 
fluctuations under heating in the $O(3)\: \sigma $ model. The main conclusion 
is that small instantons
undergo ordinary slowing down: instantons with a given size reach
the thermodinamical equilibrium distribution when the heating algorithm 
attains the corresponding length. A different conclusion in this regard  is
settled in $SU(2)$ Yang-Mills theory, where instantons are believed to undergo
a form of critical slowing down more severe than ordinary fluctuations. 
We think that this state of things is to be attributed to the pathological
characteristic of the topology in the $O(3)\: \sigma $ model, exhibiting 
ultraviolet dominance. 

As a by-product of the present work, we have obtained  a 
high-precision non-perturbative estimate
of the mixing of the lattice topological susceptibility with the 
unity operator, finding good agreement with perturbative results. This shows 
that this quantity is essentially perturbative. We have
also obtained  the size distribution function of the average 
number of instantons in the physical vacuum
in a region of sizes not yet explored, reaching sizes of a small fraction 
of the inverse mass gap, the physical length of the model. These results lay 
on a two-fold assumption:
first, instantons thermalize as ordinary fluctuations; second, instantons
do not interact among themselves. The first assumption is consistent with
other results of the present work; the second is reasonable if one takes
into account that the typical sizes of instantons involved in our analysis
are much smaller than the lattice size.

Up to the sizes explored, no physical cut in the distribution function
has been observed, so having a new evidence that the topology of the $O(3)\: 
\sigma $ model is strongly ultraviolet dominated. 
Moreover, we have found that the lattice-regularized version of the topological
susceptibility has a non-physical cut-off in the size of instantons 
corresponding to $\sim4$  lattice spacings. As a consequence, a part of the 
non-perturbative contribution to the topological susceptibility is lost 
on the lattice, at least in the region of $\beta$ up to now explored in 
numerical simulations. We attribute to this loss of non-perturbative signal
- decreasing as $\beta$ increases - the observed deviation from scaling of the 
topological susceptibility on the lattice.

We have also argued on the basis of a perturbative calculation,
that the mixing of the lattice topological susceptibility with the action 
density is a negligible part of the total non-perturbative signal.

The ultimate consideration of this work is a caveat: when the
$O(3) \: \sigma $ model is used as a laboratory to get insight into methods 
conceived for the theory of physical interest, i.e. $QCD$, 
the possibility to be wrecked in the unphysical artefacts of the model
must be seriously considered.

\vspace{1cm}
{\bf Aknowledgements}. We wish to thank Adriano Di Giacomo for having suggested
the problem, Massimiliano Ciuchini, Paolo Rossi and
Ettore Vicari for many useful and stimulating conversations.

\newpage

\begin{center}
{\bf APPENDIX}
\end{center}

\begin{center}
{\bf  Calculation of the mixing coefficient}
\end{center}

For both continuum and lattice calculations at low temperatures, the 
perturbative expansion is obtained by setting $\phi \equiv [\pi _{i}, \sqrt{1
- \sum _{i} \pi ^{2} }\; ], \; i \, = \, 1,\, 2 $. The perturbation theory
suffers from infrared divergences, which can be cured by adding a magnetic term
to the action

\beq
S_{M} \: = \:  \int d^{2}x \: h\, \sqrt{1-\sum_{i}\pi _{i}^{2}} \;\; .
\eeq
Indeed, $S_{M}$ explicitly breaks the $O(3)$ invariance and acts as a mass 
term for the $\pi $ field. The $O(3)$ invariant quantities (and the 
relations among them) are free of infrared divergences, and have a well-defined
limit for $h \rightarrow 0$.

From the standard theory of the renormalization of composite operators
 it follows that the coefficients $Z(\beta)$ and $A(\beta)$ enter
the relation  
\beq
\Gamma _{[\sum_{x}Q^{L}(x)Q^{L}(y)],\;ren}^{(n)} \; = \; a^2\, Z(\beta)^{2}\;\Gamma _{[\int_{x}Q(x)Q(y)],\;ren}^{(n)\;(\overline{MS})} \; + \; a^2\, A(\beta)\;\Gamma _{[S(y)],\;ren}^{(n)} + O(a^4) \;\;\;\;  , 
\label{eq:mast}
\eeq
where $\Gamma _{[O], \; ren}^{(n)}$ stands for $n$-point ($n\neq 0$) 
renormalized $\Gamma$-function with the insertion of the operator $O$ and 
$\overline {MS}$ is an arbitrarily chosen renormalization scheme. 
Eq.~(\ref{eq:mast}) allows to calculate $A(\beta)$, order by order in 
perturbation theory, as a series in $1/\beta $: 
$A(\beta) = \sum_n a_n/\beta ^n$.
In Eq.~(\ref{eq:mast}), an integration over the variable $y$ is allowed 
since there are no operators other than $S(x)$ with the same quantum
numbers of $\chi$ (this would be forbidden in the case of mixings with several 
operators differing  by an integration by parts). After the integration, 
Eq.~(\ref{eq:mast}) becomes
\beq
\Gamma _{[(Q^{L})^2/V],\;ren}^{(n)} \; = \;  Z(\beta)^{2}\; \Gamma _{[Q^2/V],\;ren}^{(n)\;(\overline{MS})} \; + \; A(\beta)\;\Gamma _{[S(0)],\;ren}^{(n)} + O(a^2) \;\;\; , 
\label{eq:mast1}
\eeq
where $V\, = \, L^2 a^2$ is the space-time volume.
Now the vertices of the topological charge appear in the diagrams
at zero momentum, with a considerable simplification of the calculation. 

In view of a perturbative  evaluation of $A(\beta )$, the most suitable  
choice is  $n = 2$, with an  insertion  of
two external lines of the field $\pi$ with the same index $i$.  
At the lowest order:
\beq
  \Gamma _{[S(0)],\;ren}^{(2)} = 2p^{2} \;\;\; ,
\eeq
where $p$ is 
the momentum flowing in  the external lines, and there is only one diagram 
${\cal D}$
- order $1/\beta^3 $ -
contributing to $\Gamma _{[Q^{2}/V]}^{(2)} $,
both on the lattice and on the continuum (Fig.~1). A straightforward
calculation shows that
the diagram vanishes on the continuum: ${\cal D}|_{cont} \,=\, 0$. 
As a consequence: 
\beq
{\cal D}|_{latt} \: = \: \frac{a_3}{\beta ^3 }\: 2p^2 + O(a^2) \; ;
\eeq
$a_3$ is just the coefficient we are looking for. 

We now present the complete calculation of ${\cal D}|_{latt}$.
All the results will be expressed as obtained in the infinite-size
lattice: they can be translated in the finite-size case  
by simply interpreting integrals as discrete sums. 
We first give the explicit expression of the four-legs vertex of the lattice 
topological charge operator $Q^{L}$:
\begin{equation}
\frac{1}{8\pi} \int  \Pi _{i} \frac{d^{2}p_{i}}{(2\pi )^{2}} 
\; V(p_{1}\; , \; p_{2})\; (2\pi )^{2} \delta(\; \sum _{i} p_{i}\; )
\; \pi _{1}(p_{1})\pi _{2}(p_{2})\;\pi (p_{3})\cdot\pi(p_{4}) \;\; ,
\end{equation}
where all the integrations are from $-\pi$ to $\pi$, the index $i$ runs from 1 
to 4, and 
\begin{equation}
V(p_{1}\; , \; p_{2}) \; = \; \epsilon _{\mu \nu}\;\left [ \sin p_{1\mu }\sin p_{2\nu }\; - \; \sin (p_{1}+p_{2})_{\mu}\; (\sin p_{1\nu}\; - \sin p_{2\nu}) \right] .
\end{equation}
The result of the diagram in Fig.~1 is:
\begin{equation}
\frac{1}{8\pi ^2\beta ^3}\int \frac{d^2 p_1}{(2\pi)^2}\frac{d^2 p_2}{(2\pi ) ^2
}\;\frac{{\cal I} (p\; ,p_1\; ,p_2)}{(\Box _{p_1} + h)(\Box _{p_2} + h)(\Box _{p+p_1+p_2} + h )} \;\;\; ,
\label{eq:res}
\end{equation}
where $\Box _p $ is a shorthand for the inverse propagator of the Symanzyk 
tree-level improved action 
\begin{eqnarray}
\Box _{p} \; & = &  \; \sum _{\mu}\hat{p}_\mu ^2 \; + \; \frac{1}{12} \sum_{\mu} (\hat{p}_{\mu } ^2)^2 \; \; , \\ \nonumber
\hat{p}_{\mu } ^2 \; & = &  \; \hat{p}_{\mu }\hat{p}_{\mu }^{\star} \; , \;\;\;\;\;\;\;\;\;\; \hat{p}_{\mu } \; = \; e ^{ip_{\mu}a}- 1 \;  ,
\end{eqnarray}
and
${\cal I}(p\; ,p_1\; ,p_2)$ is a compact notation for
\begin{eqnarray}
V(p\: ,p_1)V(p\: ,p_1)\;  & +   & \; V(p,p_1)V(p,p_2) \; + \; 2\, V(p,p_2)V(p_1,p_2) \; + \; \nonumber  \\
(\; V(p_1,p_2)\; ) ^2 \; &  + &  \; V(p_1,p_2)V(p+p_1+p_2\: , -p_2) .
\label{eq:vert}
\end{eqnarray}  
It may be checked that the integral in (\ref{eq:res}) is convergent in the 
limit $h\rightarrow 0$. The correct procedure would be now calculating exactly
the above integral and extracting the term proportional to $p^2$ in the result
of the integration. A look at the complete expression suffices to discourage
any attempt in this direction. The other, apparently viable, possibility of
performing Taylor expansions in the integrand and  isolating the
$p^2$ term before integration, does not  work \cite{alles}. 
We have found a way to 
get the result without any Taylor expansion, relying only upon the symmetries
of the vertex $V(p_1,p_2)$. 

Let us consider, for example, the contribution to (\ref{eq:res}) coming from
the third term in (\ref{eq:vert}):
\begin{equation}
\frac{2}{8\pi ^{2}\beta ^{3}}\int \; \frac{V(p, p_2)V(p_1,p_2)}{(\Box _{p_1} + h)(\Box _{p_2} + h)(\Box _{p+p_1+p_2} + h )}\; \equiv \; \frac{1}{8\pi ^{2}\beta ^{3}}{\cal I} _{3} .
\end{equation}
After the translation $p_1 \rightarrow p_1 - p $ we have
\begin{equation}
{\cal I} _{3}\; = \; 2\int\; \frac{V(p,\; p_2)V(p_1-p\;,\; p_2)}{(\Box _{p_1-p} + h)(\Box _{p_2} + h)(\Box _{p_1+p_2} + h )} .
\end{equation}
We observe that $V(p,p_{2})$ factors out a $\sin p \sim p $ term, while 
$V(p_1-p\:,\, p_2)$ may be decomposed into $V(p_1,p_2)$ and the remaining part,
$V_r(p,p_1,p_2)$, which is $O(p)$. When $V(p,p_2)$ multiplies $V_r$ the
$p^2$ term comes out trivially and it is sufficient to put $p = 0$ in the
propagators; the part with $V(p,p_2)V(p_1,p_2)$ reduces to $-{\cal I}_{3}$ 
after the following transformation of the integration variable $p_1$: 
$p_1 \rightarrow p_1-p_2$, \ \ $p_1 \rightarrow -\; p_1$ 
and using the property $V(-p_1-p_2\;,\, p_2)=-\; V(p_1,p_2)$. 

A similar procedure works for the contribution to (\ref{eq:res}) from
the fourth and the fifth terms in (\ref{eq:vert}), while the first and the
second terms need no manipulation because the factorization of $p^2$ is
trivial.

The sum of the various terms in (\ref{eq:res}) after the factorization of 
$2p^{2}$ is an IR-convergent integral where $h$ can be safely put equal to 
zero; the mixing coefficient $ a_3$ we are looking for, is:  
\begin{eqnarray}
&\frac{1}{16\pi ^2} \int  \; \Box _{p_1}^{-1}\Box _{p_2}^{-1}
\Box _{p_1+p_2}^{-1} \left\{ \sin ^2 p_{1\mu }(2 + \cos p_{1\nu })^2
+ \sin p_{1\mu}\sin p_{2\mu}(2 + \cos p_{1\nu })(2 \right. &+ \cos p_{2\nu }) 
\;\;
\nonumber \\
&\sin  p_{2\nu }(2+\cos p_{2\mu })[\;(\sin p_{1\nu }-\sin p_{2\nu})
\cos (p_1+p_2)_{\mu}-(\sin p_{2\nu }-\sin (p_1+p_2 &)_{\nu})\cos p_{1\mu}\;]  
\nonumber \\
&\left. + \frac{1}{2}\;[\; (\sin p_{1\nu }-\sin p_{2\nu})
\cos (p_1+p_2)_{\mu} - (\sin p_{2\nu }- \sin (p_1+p_2)_{\nu})\cos p_{1\mu}\;]^2
\right\} \;\; , &
\end{eqnarray} 
where there is no sum over repeated indices; $\mu $ and $\nu$ indicate 
two different directions.

In Table I we report values of $a_3$ for various lattices.

In order to evaluate the relevance of the mixing of $\chi^L$ with the action 
density in the
determination of $\chi$ from lattice simulations, we now need  an estimate
of the condensate $\langle S(0) \rangle $. The only 
available result comes from the large $N$ limit \cite{roca}:
\beq
\langle S(0) \rangle |_{N\rightarrow\infty} \: = \:
-\frac{1}{2N}m^2 \:\:\; .
\eeq
By using the exact relation between the mass gap of the theory, $m$, and 
the scale invariant parameter of the
lattice Symanzik-improved  theory, $\Lambda _{SY}$ \cite{hmn}, 
we obtain the estimate for $N = 3$ 
\beq   
\langle S(0) \rangle \: \simeq \:
-200\, \Lambda_{SY}^{2} \: \: \: .
\eeq
This result, combined with our estimate for $A(\beta)$, gives a 
mixing term which is negligible with respect to the topological
non-perturbative signal: at $\beta = 1.45$, for instance, 
$A(\beta)\langle S(0)\rangle/[ Z^2(\beta)\chi ] \sim 6\cdot 10^{-4} $.
We believe that higher order 
corrections in $1/N$ and $1/\beta$ cannot spoil the validity of this argument.

\newpage


\noindent
TABLE I: $a_3$ for a variety of lattices. $a_3$ is the first non-vanishing
term in the perturbative expansion $A(\beta) = \sum_n a_n/\beta ^n$
of the mixing coefficient of the topological susceptibility with the 
action density. $L$ is the lattice size.
\newline

\begin{tabular}{c||c|c||c|c||c|c||c}     \cline{2 - 7}
 & $ \;\;\;\; L \;\;\;\; $ & $ \;\;\; a_3\times 10^3 \;\;\; $  & $ \;\;\;\; L \;\;\;\; $ &
 $ \;\;\; a_3\times 10^3 \;\;\; $  & $ \;\;\;\; L \;\;\;\; $ &  $ \;\;\; a_3\times 10^3 \;\;\; $ & \\
 \cline{2 - 7}
    &   10    & 0.31450626  &   60    & 0.33496730  &  110   & 0.33538507 &  \\
    &   20    & 0.33022955  &   70    & 0.33512505  &  120   & 0.33541334 &  \\
    &   30    & 0.33318636  &   80    & 0.33522745  &  130   & 0.33543534 &  \\
    &   40    & 0.33422462  &   90    & 0.33529767  &  140   & 0.33545280 &  \\
    &   50    & 0.33470577  &   100   & 0.33534790  &$\infty$& 0.33556212 &  \\
\cline{2 - 7}
\end{tabular}

\vspace{2.cm}


\noindent
Table II:
$Z(\beta)$ versus $\beta$. $Z(\beta)_{2loops}$ is the multiplicative
renormalization calculated to two loops.  
$Z(\beta)_{MC}$ is the multiplicative renormalization
calculated by heating an instanton; the size of the lattice is $120$; 
Stat is the statistic of the simulation. Since data on the plateau are
correlated, we report the average error of data. 
\newline

\begin{tabular}{c||c|c|c|c||c} \cline{2 - 5}  
$\;\;\;\;\;\;\;\;\;\;\;\;\;\;\;\;\;\;\; $ & $\;\;\;\;\;\beta\;\;\;\;\;$ & $\;\;\;\; Z(\beta)_{2loops}\;\;\;\; $  & $\;\;\;\;Z(\beta)_{MC}\;\;\;\; $ & $\;\;$ Stat $\;\;$ & $\;\;\;\;\;\;\; $ \\  \cline{2 - 5}  
&  2.         &  0.643                   &  0.6285(26)       &  16000 &  \\
&  1.65       &  0.564                   &  0.5298(42)       &  16000 &  \\
&  1.45       &  0.500                   &  0.4362(31)       &  64000 &  \\ 
  \cline{2 - 5}  
\end{tabular}

\newpage

\noindent
Table III:
$P(\beta)$ versus $\beta$ for different choices of the threshold. 
$P(\beta)_{pert}$ includes the perturbative
tail calculated to four loops and   higher order terms
fitted from thermal equilibrium Monte Carlo data at large $\beta$. 
$P(\beta)_{np}$ is the perturbative tail
calculated by Monte Carlo techniques; the size of the lattice is $120$; 
Stat is the statistic of the simulation. 
\newline

\begin{tabular}{c||c|c|c|c|c||c}    \cline{2 - 6}
& $\;\;\;\;\;\;\beta\;\;\;\;\;\;$   &   Threshold   &  $\;\;\;P(\beta)_{pert}\times 10^{5}\;\;\;$  &  $\;\;\;P(\beta)_{np}\times 10^{5}\;\;\;$  & $\;\;\;$ Stat $\;\;\; $ & \\ \cline{2 - 6}
&         &    0.2     &               &    5.24(6)     &       &  \\
&  1.45   &    0.4     &    5.21(14)   &    5.26(5)     &  5000 &  \\
&         &    0.6     &               &    5.28(6)     &       &  \\
\cline{2 - 6} 
&         &    0.2     &               &    4.22(3)     &       &  \\
&  1.50   &    0.4     &    4.20(12)   &    4.23(3)     &  5000 &  \\
&         &    0.6     &               &    4.24(3)     &       &  \\ 
\cline{2 - 6} 
&         &    0.2     &               &    3.82(2)     &       &  \\
&  1.525  &    0.4     &    3.89(10)   &    3.83(2)     &  5000 &  \\
&         &    0.6     &               &    3.86(2)     &       &  \\  
\cline{2 - 6}
&         &    0.2     &               &    3.50(2)     &       &  \\
&  1.55   &    0.4     &    3.54(9)    &    3.51(2)     &  5000 &  \\   
&         &    0.6     &               &    3.53(2)     &       &  \\    
\cline{2 - 6}
\end{tabular}

\newpage
\begin{center}
FIGURES
\end{center}

\vspace{1 cm}

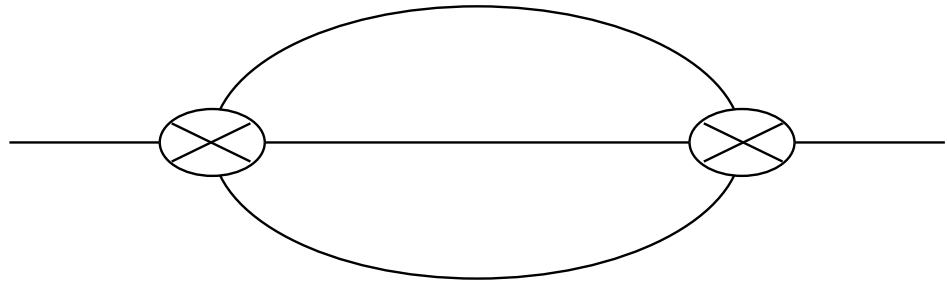
FIG. 1. Diagram contributing to the two-point proper
function with insertion of $(Q^L)^2/V$. White crossed blobs
indicate the charge operator.

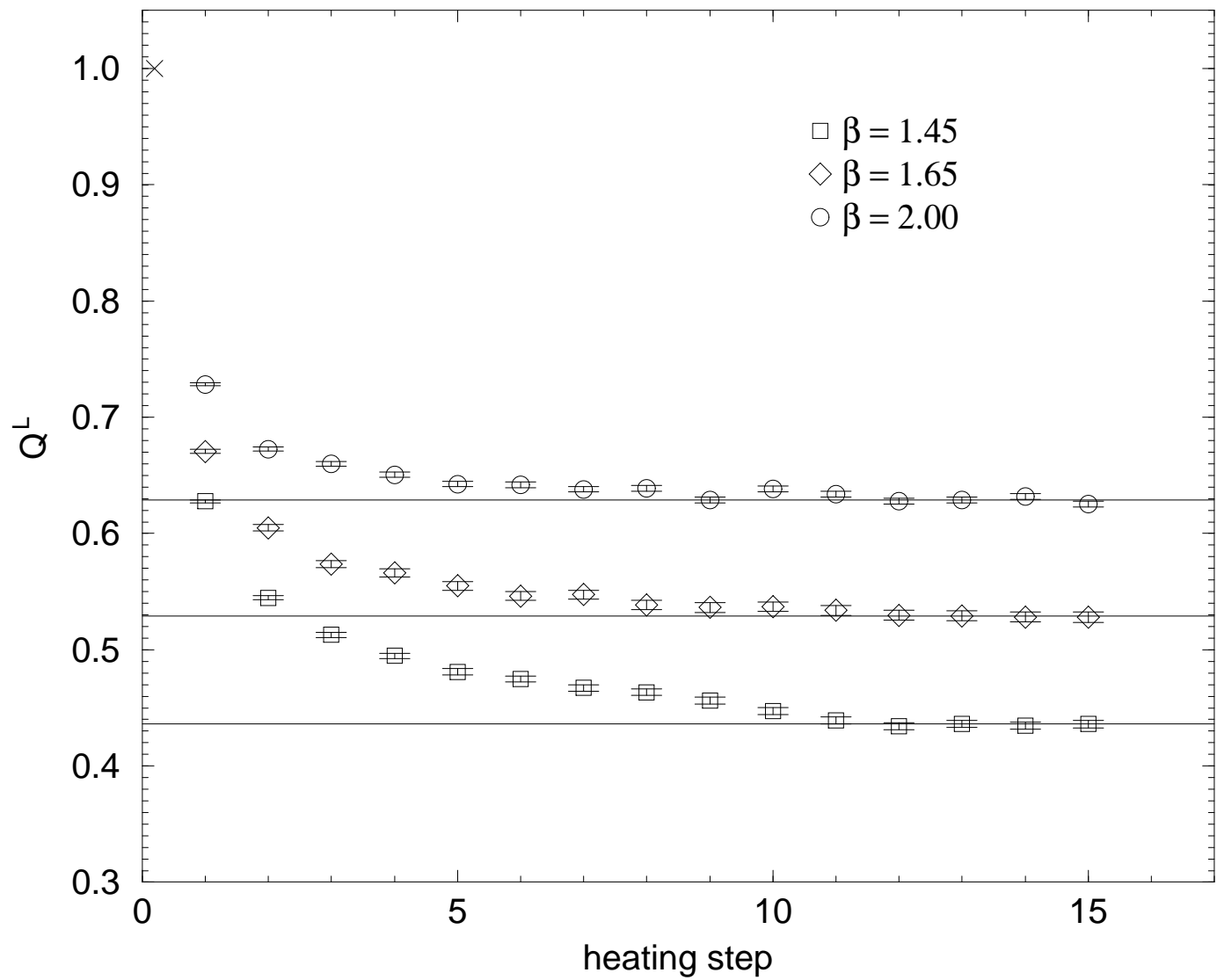
FIG. 2. $Q^{L}$ versus the heating step. 
The thermalization is performed by the heat-bath
algorithm at $\beta = 1.45 $, $1.65$, $2.00$  on a $120^2$ lattice. 
The solid lines indicate the
values of $Z(\beta)$ estimated by averaging data at the plateaux.

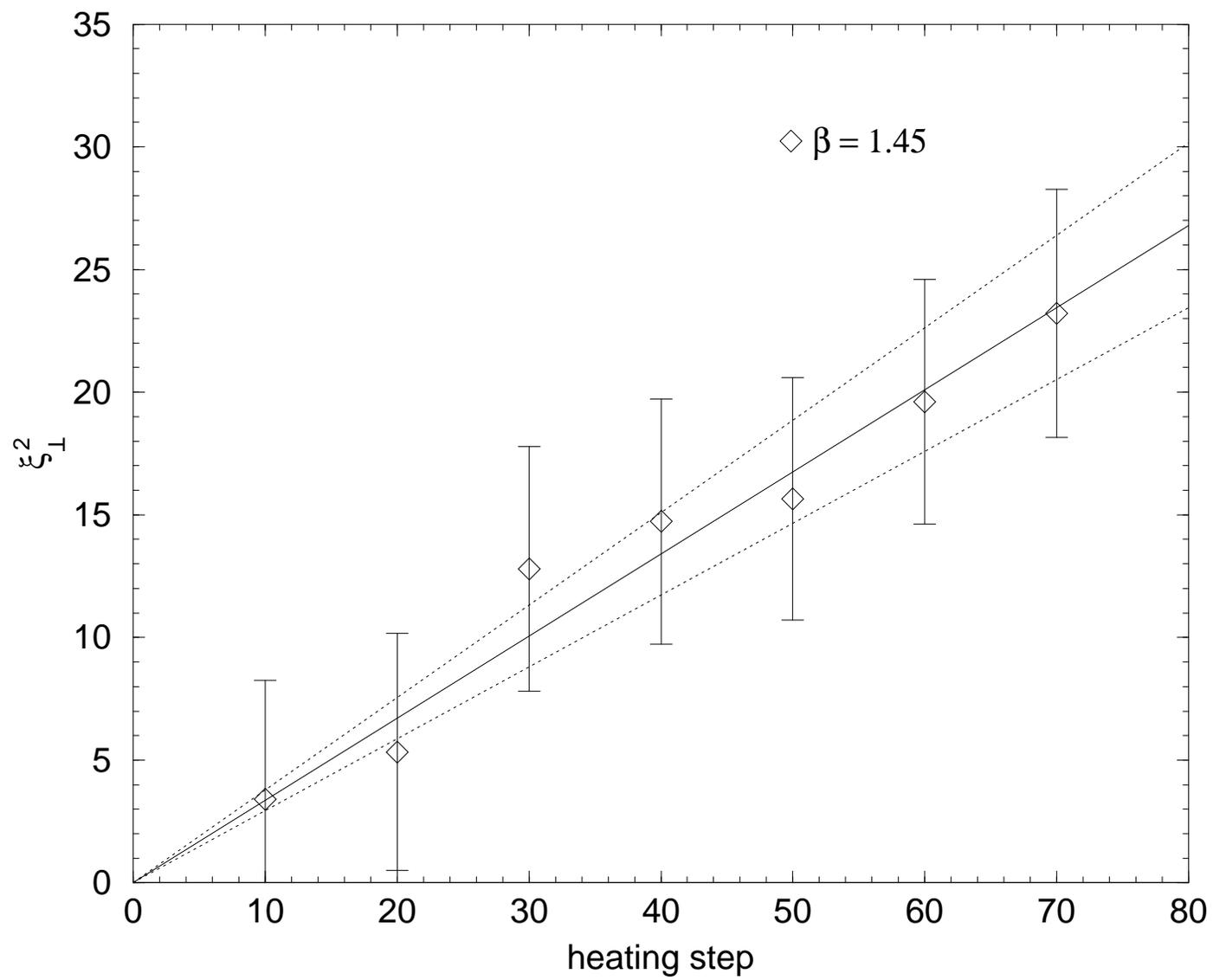
FIG. 3. $\xi^{2}_{\perp}$  versus the heating step with the heat-bath algorithm
at $\beta = 1.45$. The solid line indicates the best fit with the law 
$\xi^{2}_{\perp}\, =\, c\, t$ (with the relative error).

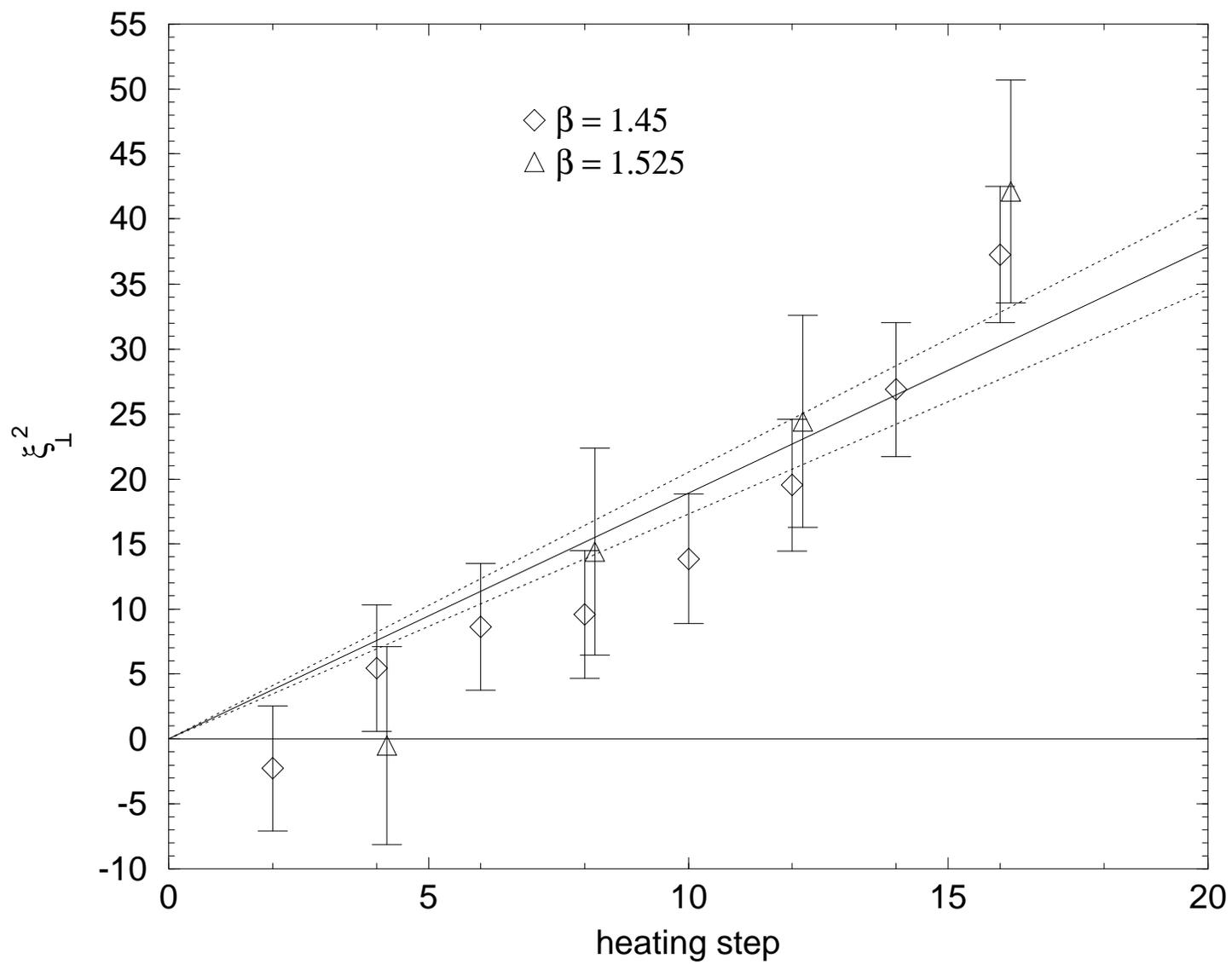
FIG. 4. As in Fig. 5 with the overheat-bath algorithm. Data are 
obtained at two different values of $\beta$: $1.45$ and $1.525$.

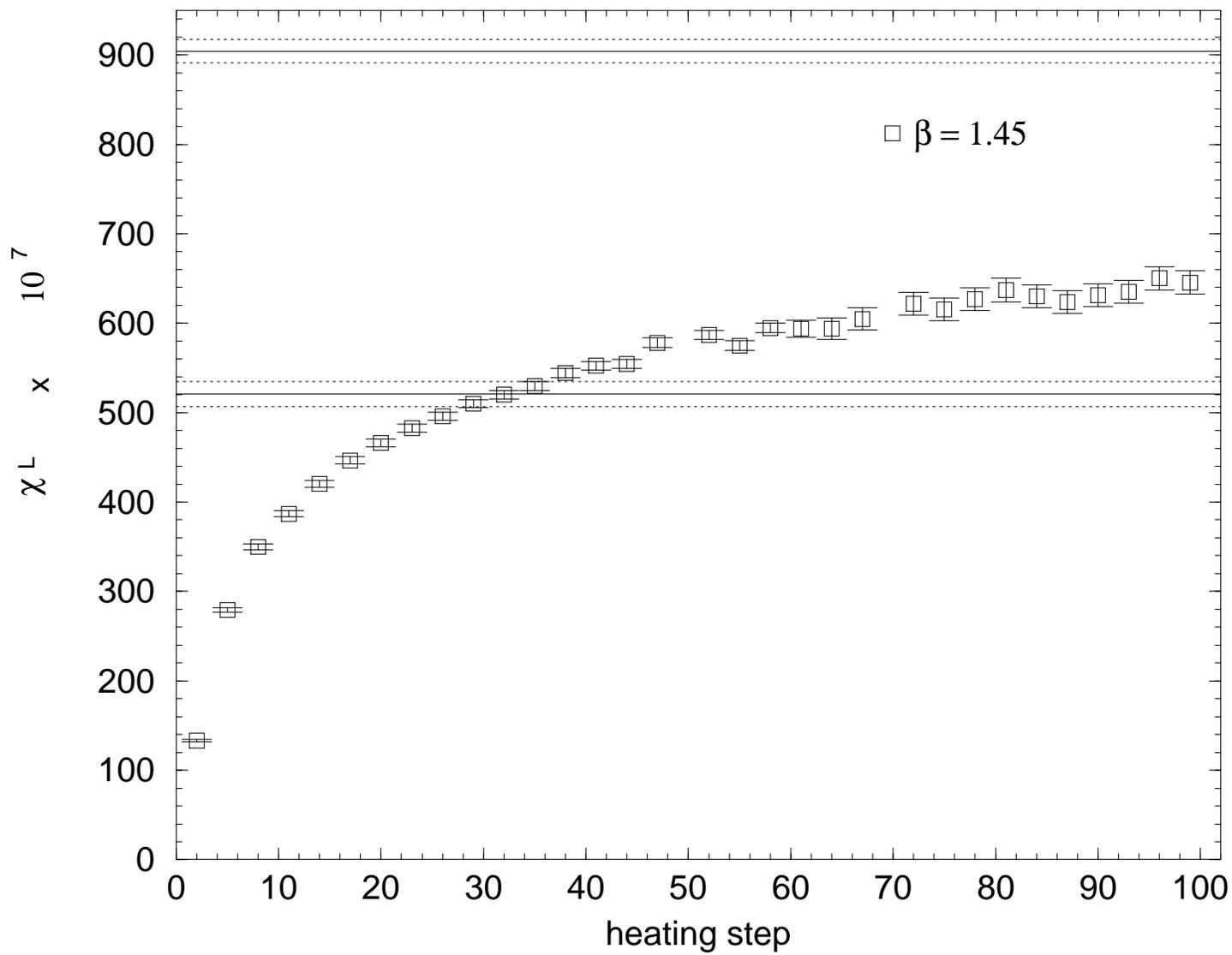
FIG. 5. $\chi ^L$ versus the heating step at $\beta = 1.45$ starting from the
flat configuration on a $120^2$ lattice. The thermalization is performed by the
heat-bath algorithm. The two solid lines indicate the estimate of $P(\beta)$
by perturbative calculation and the value of $\chi^{L}$ obtained at the thermal
equilibium (with the respective errors).

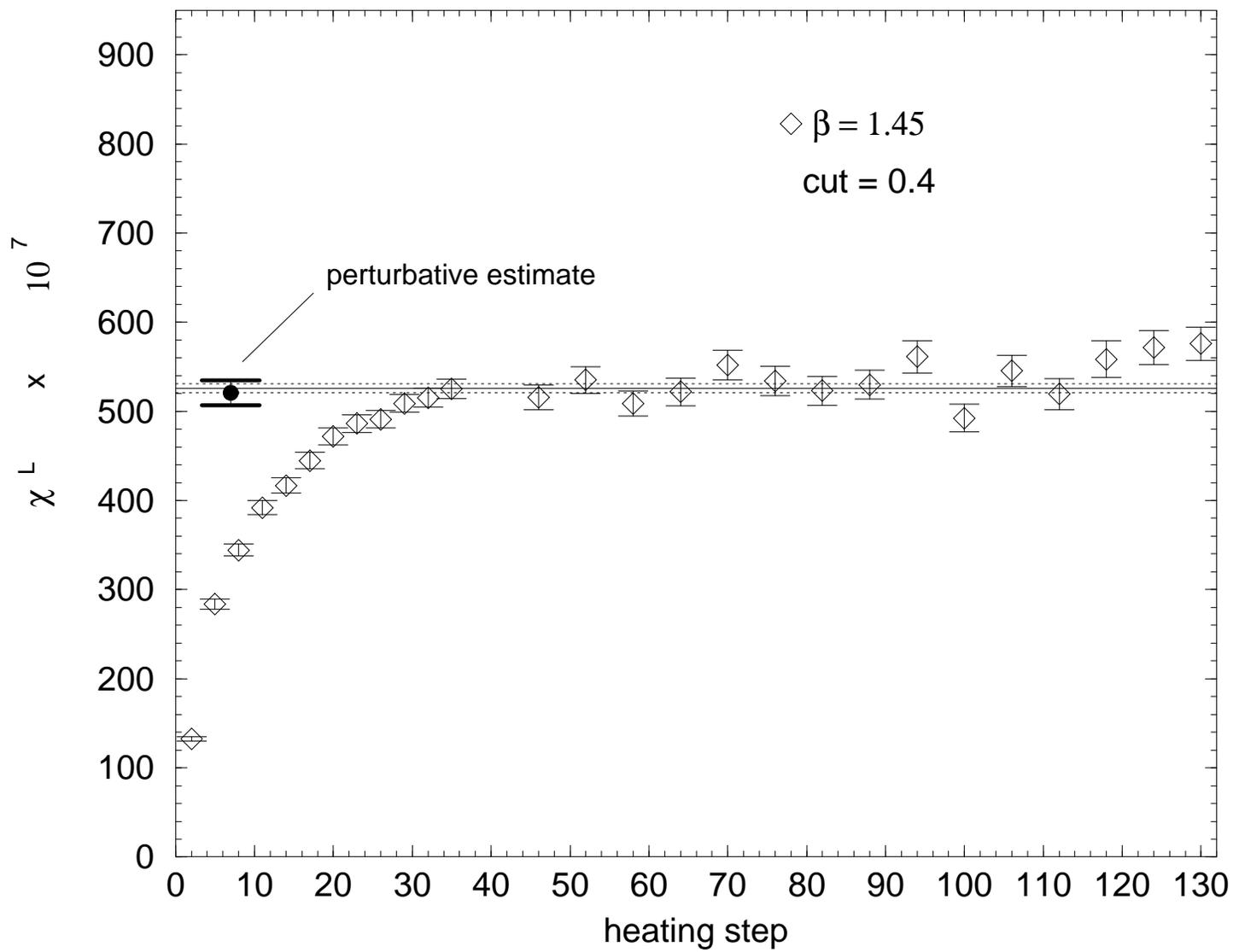
FIG. 6. $\chi ^L$ versus the heating step  at $\beta = 1.45$ starting from the 
flat configuration on a $120^2$ lattice. The thermalization is performed by the
heat-bath algorithm with the procedure of subtraction of topological
configurations described in the text. The threshold is $0.4$.
The solid line indicates the value of $P(\beta)$ estimated averaging data 
at the plateau (with the relative error).

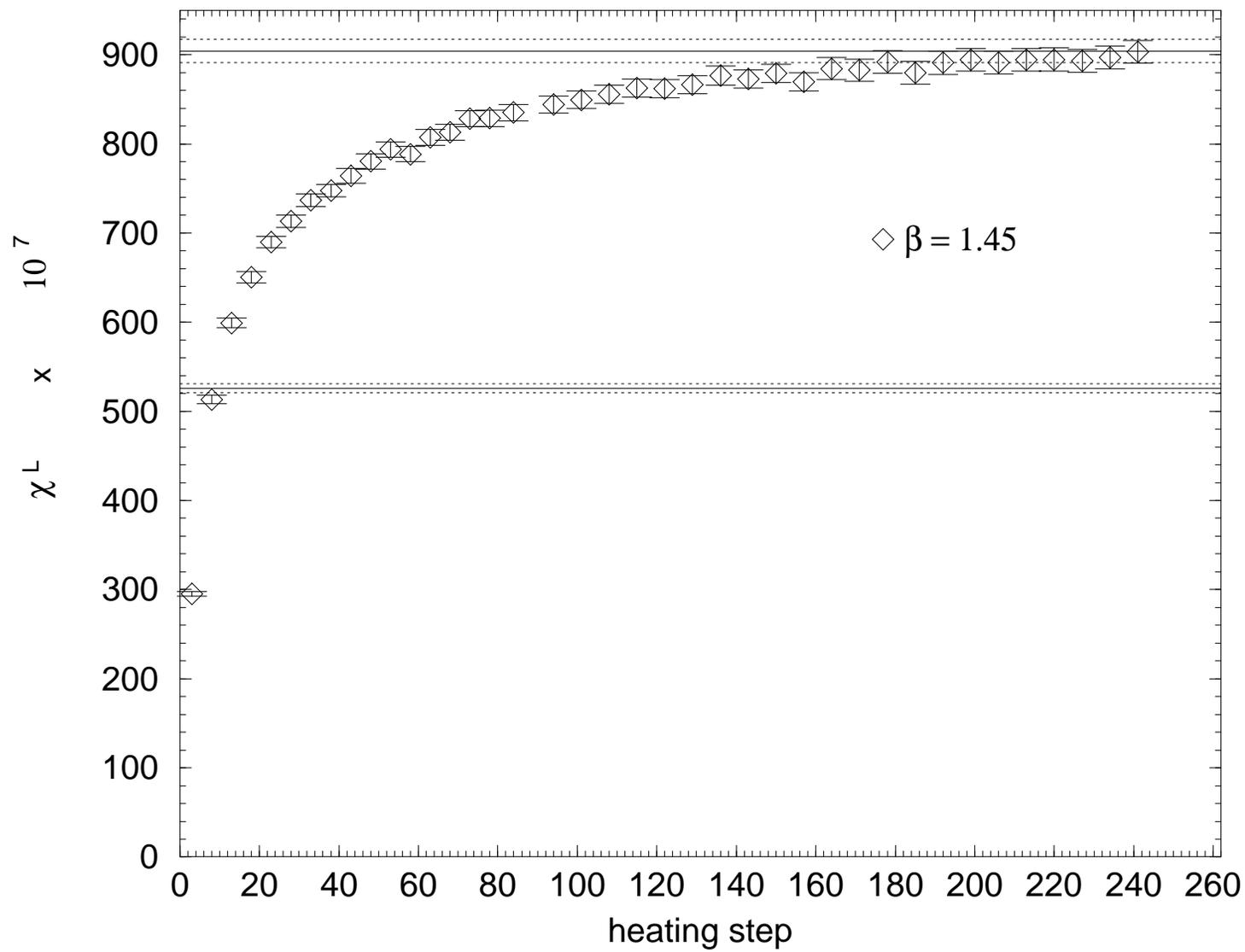
FIG. 7. The approach to equilibrium of $\chi^L$ at $\beta = 1.45$
with the overheat-bath algorithm. The two solid lines are 
the non-perturbative estimate of $P(\beta)$ and 
the value of $\chi^L$  obtained by standard MC simulations 
at the thermal equilibrium (with the respective errors).

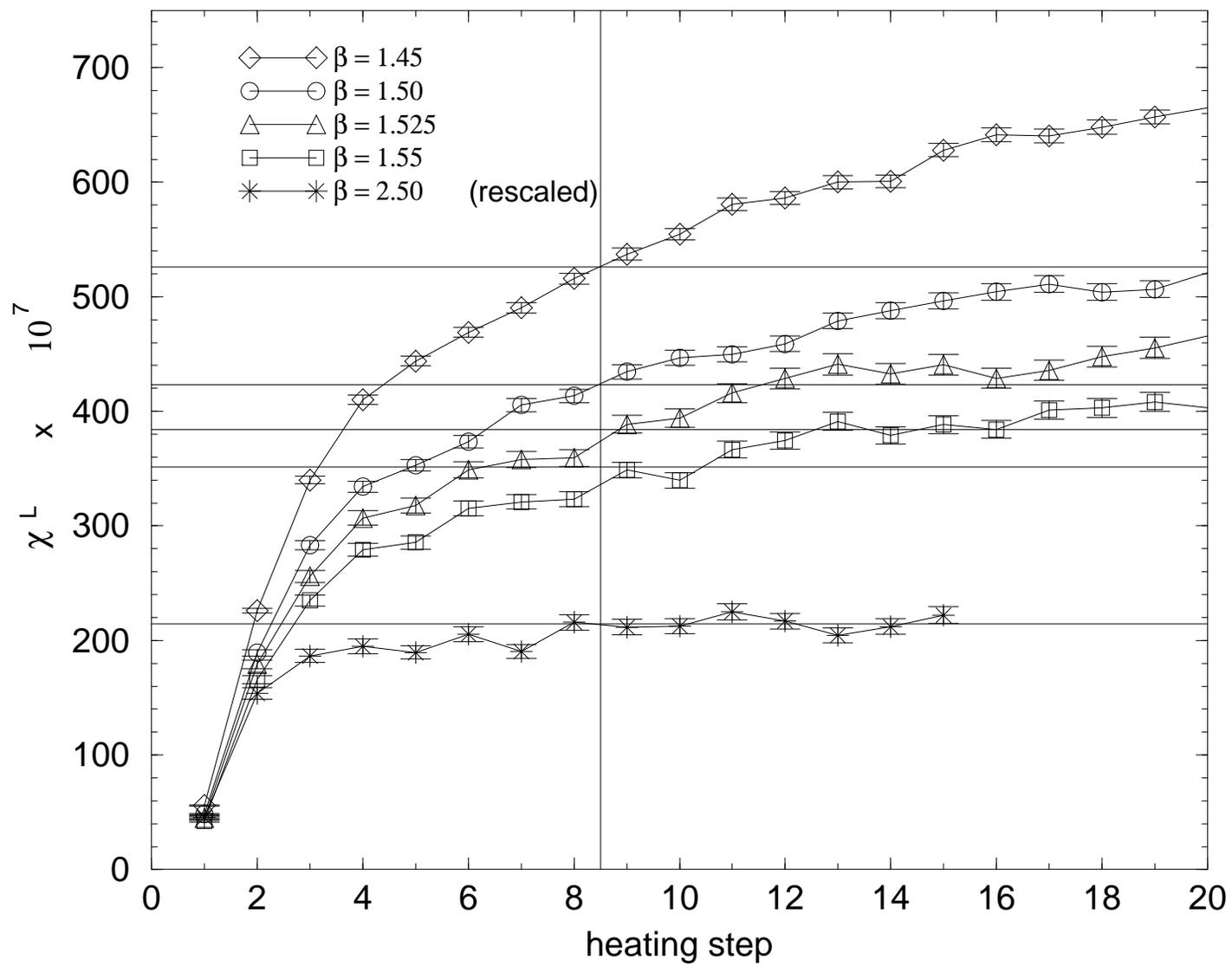
FIG. 8. The overcoming of the perturbative value
during the heating of $\chi^L$ at various $\beta$. The solid lines 
are the perturbative values. 

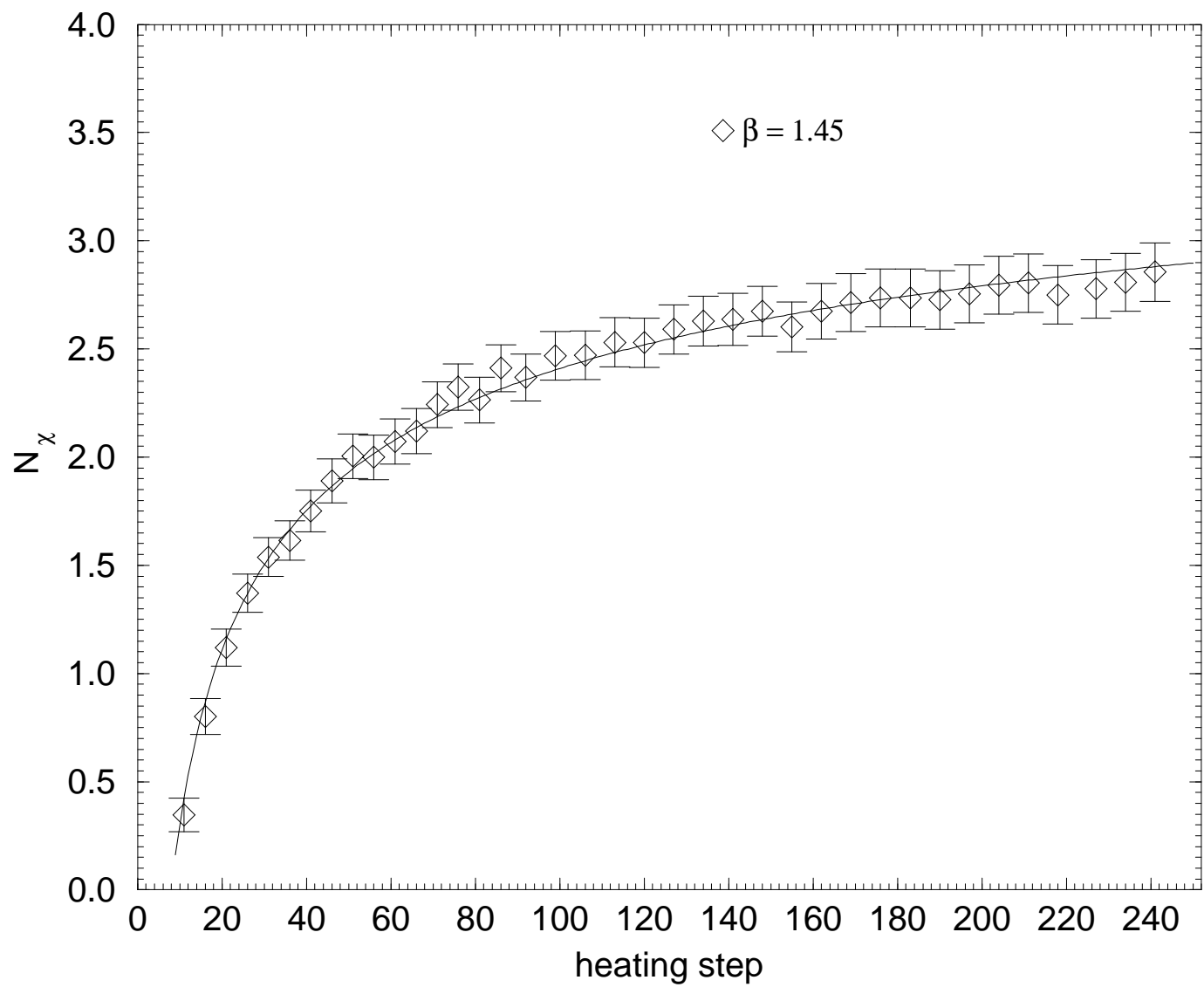
FIG. 9. The average number of instantons obtained  from 
$\chi^{L}$ at $\beta = 1.45$ versus the heating step. The solid line
represents the predicted behavior in our model of thermalization 
(see the text). 

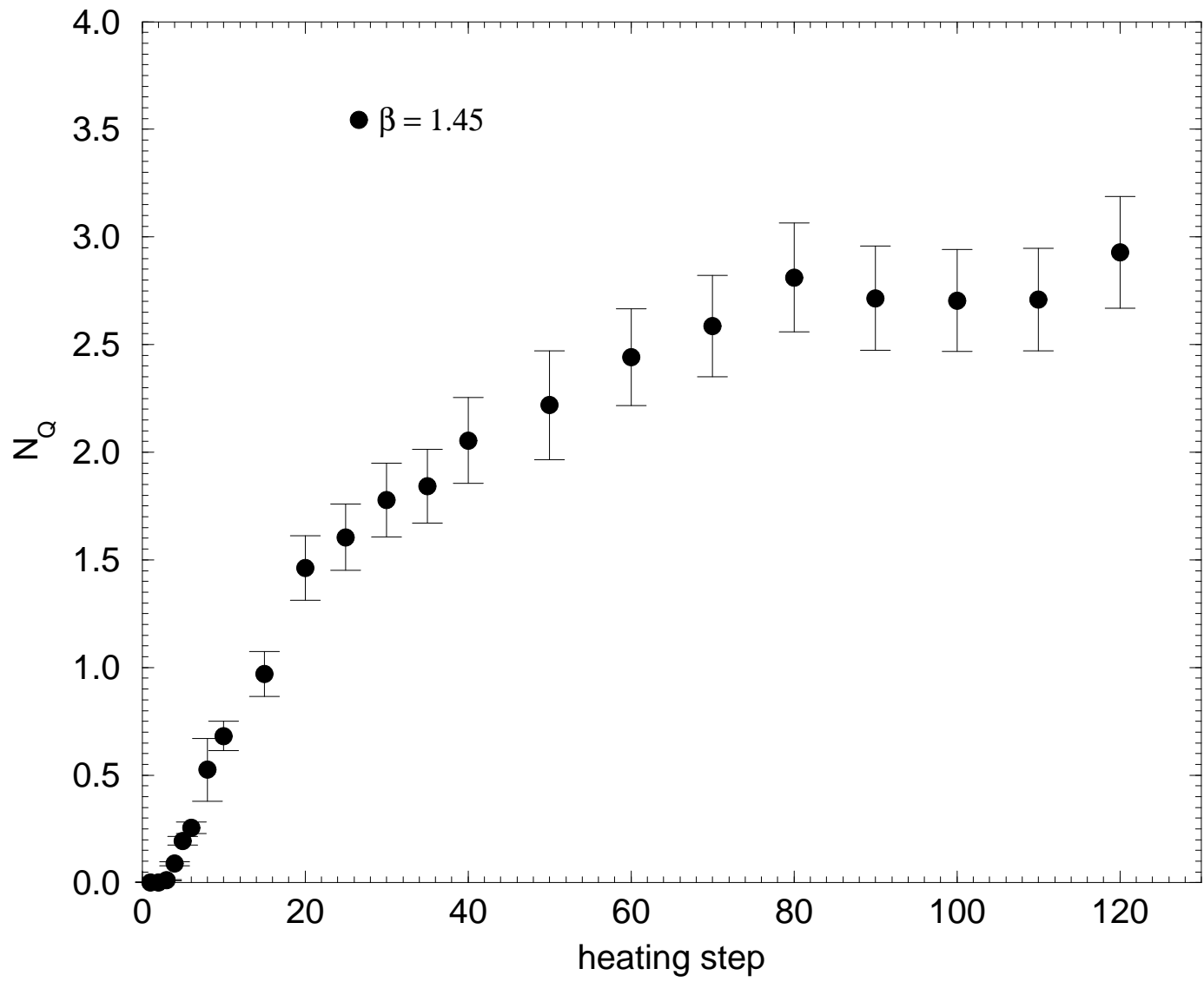
FIG. 10. The same quantity of Fig. 9 obtained from 
the frequency of the configurations in the various topological sectors.

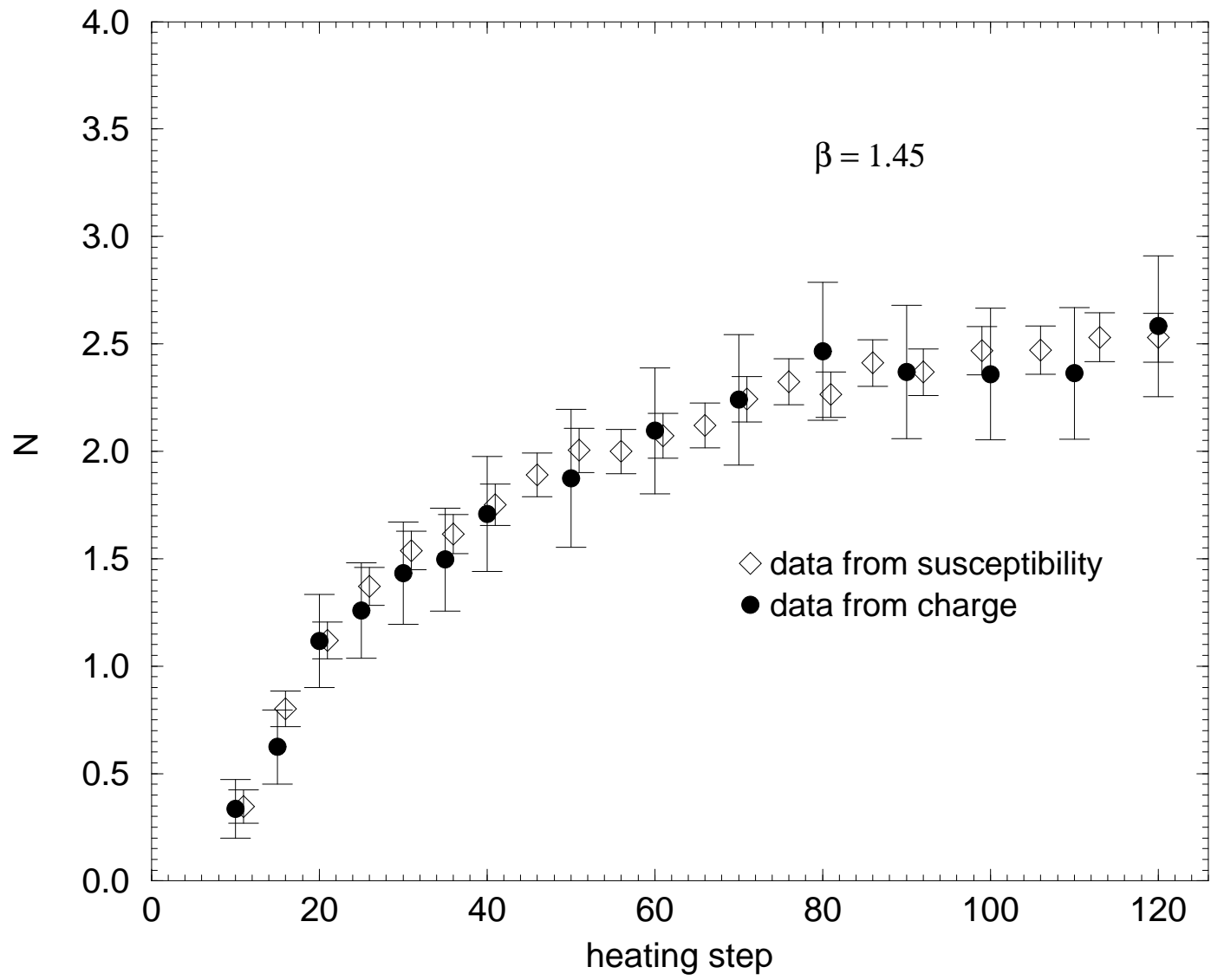
FIG. 11. Comparison of the quantities of Figs. 9 and 10
after the subtraction from data of Fig. 10
of the instantons with sizes smaller than
the the typical perturbative ranges, which give no
contribution to the non-perturbative part of $\chi^L$.

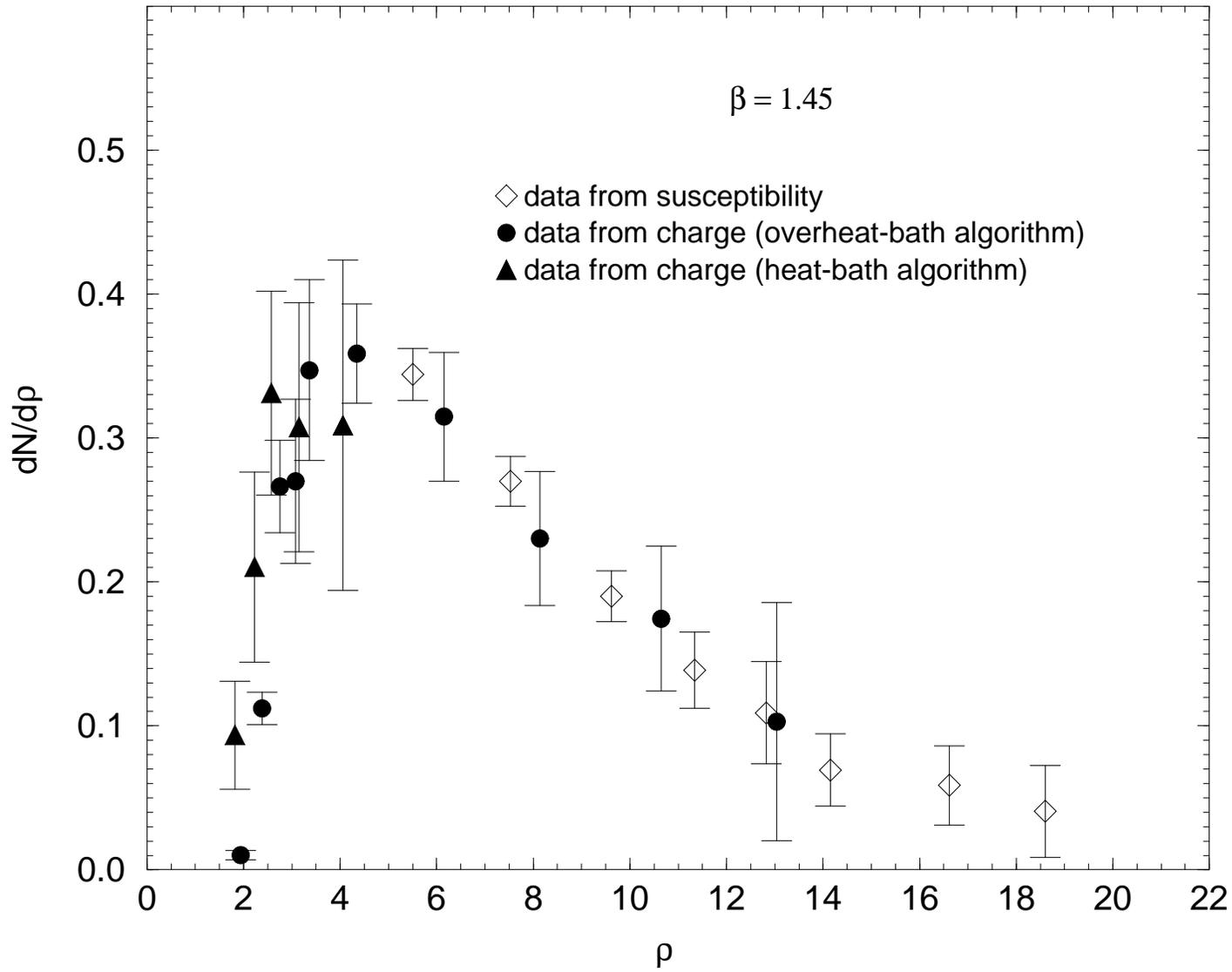
FIG. 12. The instanton size distribution (in lattice units)
as obtained from
the two methods described in the text, at $\beta = 1.45$.

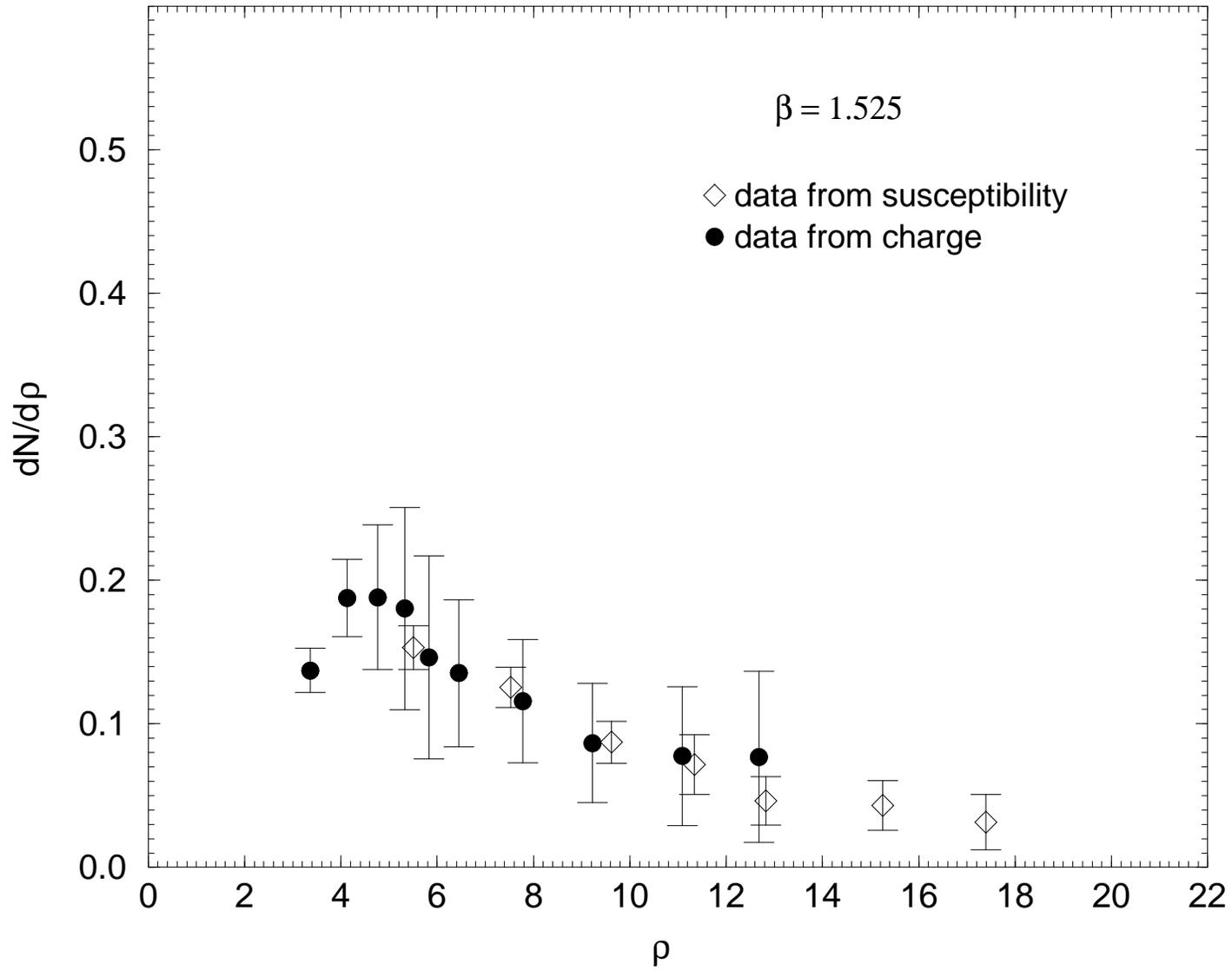
FIG. 13. As in Fig. 12 with $\beta = 1.525$.

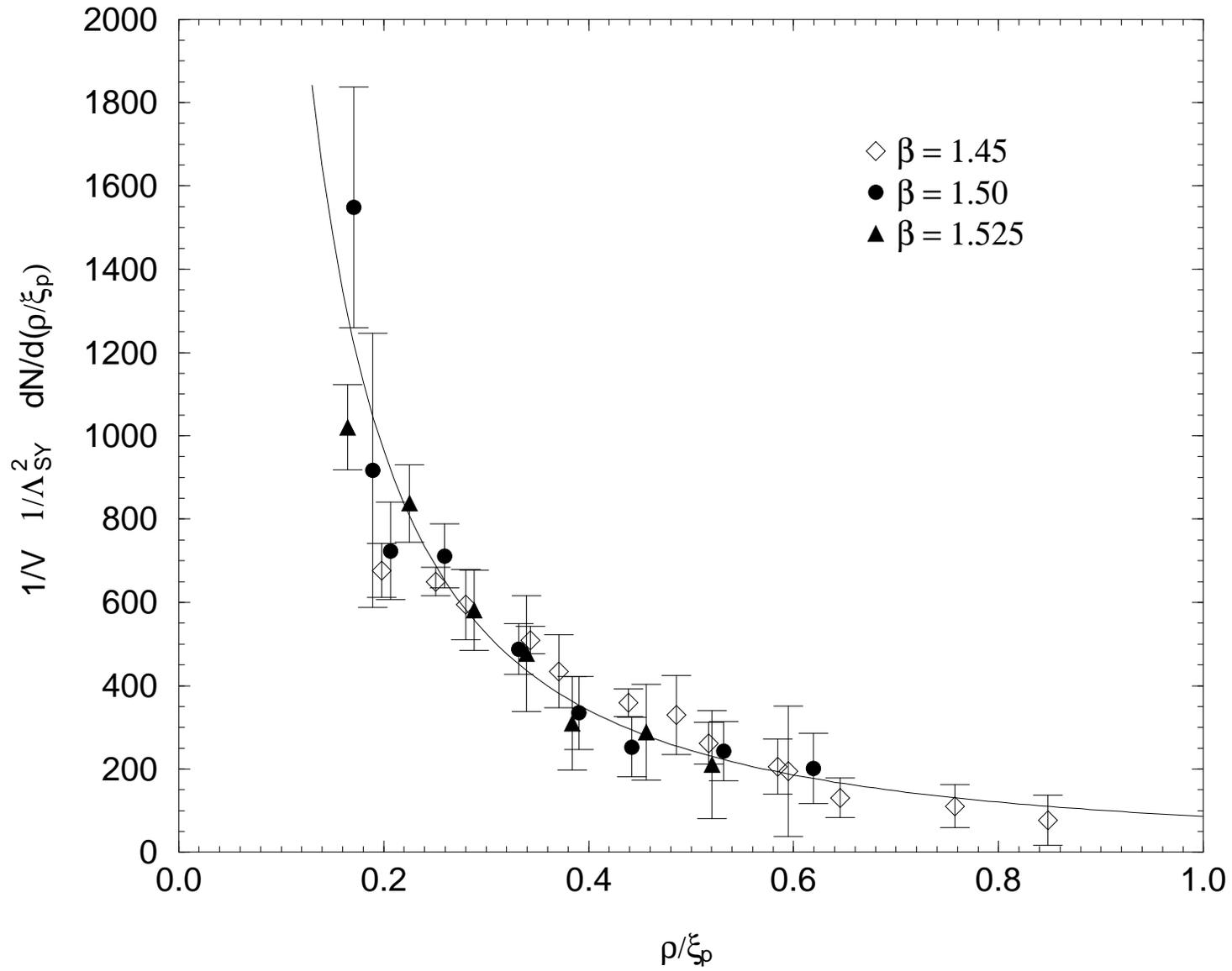
FIG. 14. The instanton size distribution in physical units 
obtained collecting data at different values of $\beta$. The solid 
line is the best fit with $n = 1.5$.

\end{document}